\documentstyle[twocolumn,prl,aps]{revtex}

\begin{document}

\twocolumn[\hsize\textwidth\columnwidth\hsize\csname @twocolumnfalse\endcsname
\draft 
\preprint{} 
\title{Phase Diagram of Integer Quantum Hall Effect} 
\author{D.N. Sheng and Z.Y. Weng} 
\address{Texas Center for Superconductivity, 
University of Houston, Houston, TX 77204-5506}  
\maketitle
\date{today}
\begin{abstract} 
The phase diagram of integer quantum Hall effect is numerically determined 
in the tight-binding model, which can account for overall features of 
recently obtained experimental phase diagram with direct transitions from
high plateau states to the insulator. In particular, the quantum 
Hall plateaus are terminated by two distinct insulating regimes, 
characterized by the Hall resistance with classic and quantized values, 
respectively, which is also in good agreement with experiments. The physical
origin of this peculiar phase diagram is discussed.

\end{abstract}
\pacs{73.40.Hm, 71.30.+h, 73.20.Jc}]

The previously proposed\cite{phased} theoretical global phase diagram 
(GPD) of the quantum Hall (QH) effect predicts that only the $\nu =1$  
QH liquid state is adjacent to the insulator, while higher QH plateau states
do not neighbor with the insulator as schematically illustrated in 
Fig. 1(a). As a result, a direct transition from 
higher QH plateau ($\nu >1$) states to the insulator is prohibited. 
However, a series of recent experimental 
measurements\cite {krav,song,others} 
have indicated a phase diagram with qualitatively different topology. 
Most recently, it has been established experimentally by Hilke {\it et 
al.} \cite{hilke1} that direct transitions 
from $\nu=1,2,...,7$ to the insulating phase can all take place as clearly
shown in Fig. 1(b). These experiments have challenged the basic theoretical
understanding of QH systems\cite{qhe} at weak magnetic field\cite{laughlin}.

These QH liquid-insulator transitions have also exhibited distinct 
properties in different regimes. For example, the Hall resistance $\rho _{xy}$ 
remains quantized at $h/e^2$\cite{hilke2,shahar} in the transition region 
between the $\nu =1$ QH state and the insulator at strong magnetic field or low 
Landau-level (LL) filling factor $n_{\nu}$, even though the longitudinal 
resistance $\rho_{xx}$ already increases almost by one order of magnitude above 
the value $h/e^2$ in the same region denoted as II in Fig. 1(b). By contrast, 
at weak magnetic field or higher $n_{\nu}$, $\rho _{xy}$ near the critical 
region of the $\nu>1$ QH states to insulator transition becomes 
$n_{\nu}$-dependent\cite{song} and is very close to the classic value $B/nec$ 
rather than the quantized value. In fact, such a classic behavior of 
$\rho_{xy}$ has been found\cite{krav,hilke2} to persist well into 
an insulating regime designated by I in Fig, 1(b), which suggests that there 
are two distinct insulating regimes surrounding the QH liquid states as 
opposed to the single one\cite{phased}.

Direct transitions between the $\nu >1$ QH states and the insulator have been 
already found in the tight-binding model (TBM) based on numerical 
calculations\cite{dns,xie,others1}. But the relevance of such a lattice model 
to the experiment is still controversial\cite{bhatt}, since the strength of the 
magnetic field usually cannot be reduced weak enough to directly simulate the 
realistic situation within the numerical capacity. Therefore, it is 
particularly important to identify the overall phase diagram and corresponding 
transport properties in such a model in order to understand the underlying 
physics and establish a real connection with the experiments.

In this Letter, we obtain, for the first time, a numerical ``global phase 
diagram'' for the integer QH effect based on the TBM and the results are 
summarized in Fig. 1(c).  The topology of the phase diagram is strikingly
similar to the experimental one shown in Fig. 1(b), and in particular the 
insulating phase is indeed divided into two regimes: In a strong disorder and 
low magnetic field region (Insulator I), the Hall resistance follows a classic 
value while the longitudinal resistance shows insulating behavior; in a weak 
disorder and high magnetic field region (Insulator II), we find that 
$\rho _{xy}$ remains at the quantized value $h/e^2$ near the transition region 
even when $\rho _{xx}$ increases up to $8h/e^2$. Both are in good agreement 
with the aforementioned experiments.  Finally we provide a physical 
interpretation of the nature of the present non-float-up phase diagram based 
on the calculation of the equilibrium edge current. 

The phase diagram in Fig. 1(c) can be determined by following the trace of 
extended levels by continuously tuning the disorder strength or magnetic 
field $B$. The position of each extended level forms a boundary which separates 
a given QH plateau state from another QH state or the insulating phase, and 
can be identified by the peak (which is sample size independent)
 of the density of states carrying nonzero Chern 
number\cite{dns,bhatt}, calculated based on the TBM Hamiltonian 
$H=-\sum_{<ij>}e^{ia_{ij}}c_i^{+}c_j+H.c.+ \sum_iw_ic_i^{+}c_i$, which is 
characterized by two parameters: the magnetic flux per plaquette
$\phi =\sum_{\Box }a_{ij}=2\pi /M$ and the disorder strength $W$ of the random 
potential $w_i$: $|w_i|<W/2$. The result shown in Fig. 1(c) is calculated at 
$M=64$. In the Chern number calculation, the sample size is up to 
$64\times 64$.  The position of an extended level can be equally identified by 
the peak of the longitudinal conductance $\sigma_{xx}$, which coincides with  
the Chern number result, but this latter method has an advantage as it can be 
applied to much weaker magnetic fields. At $M=384$, the sample size in 
calculating $\sigma_{xx}$ is up to $L_x=200$ and $L_y=10^3 M$ using transfer
matrix method\cite{land}.
The phase diagram in Fig. 1(c) remains essentially the same as we continuously 
change the magnetic flux from $M=8$ to $384$. Note that $W_c$ 
(which depends on the magnetic field B) is the critical value at which the 
last QH plateau state disappears and the system becomes an insulator.

The similarity between the numerical phase diagram [Fig. 1(c)] and the
experimental one [Fig. 1(b)] is striking. 
Similar experimental phase diagram is also obtained earlier in 
Ref.\onlinecite{krav}. Several detailed features in Fig. 1(c) are worth 
mentioning. Firstly, starting from the strong-magnetic-field insulator II and 
reducing B continuously at a fixed electron density, we obtain numerically a 
dashed curve $A$ shown in Fig. 1(c) which cuts through different phases with a 
transition pattern $0-1-2-3-0$. [As B is reduced, $W/W_c(B)$ increases due to 
the $B$-dependence of $W_c$].  Such a scan curve should be equivalent to a 
constant gate voltage $V_G$ line in Fig. 1(b) as the fixed $V_G$ means both the 
disorder strength and electron density are constants. Secondly, one can clearly 
see that all the higher extended levels as boundaries separating different QH 
states are almost vertical lines in Fig. 1(c) which do not ``float up'' much 
in terms of the LL filling number $n_{\nu} $ 
at increasing disorder strength. The same non-float-up picture also 
unequivocally shows in the experimental phase diagram of Fig. 1(b). Only the 
lowest one which defines the boundary between $\nu =1$ and the insulator floats 
up, also in agreement with an earlier experiment\cite{jiang}. 

The scan curve A in Fig. 1(c) connects insulating regimes at two ends. Let
us first focus on the insulating region I which has a boundary neighboring
with the high-plateau QH states. Fig. 2 shows the calculated $\rho _{xx}$
and $\rho _{xy}$ versus $n_\nu $ at fixed $W$ [parallel to the scan
line B illustrated in Fig. 1(c)]. The prominent feature in this region is
that $\rho _{xy}$ follows the classic behavior (the dashed curve in Fig. 2
denotes $\frac 1{n_\nu }\frac h{e^2}=\frac B{nec}$): In Fig. 2(a), the
magnetic filed is fixed at $M=32$ while different disorder strengths are
considered. Even though $\rho _{xx}$ in the inset grows with $W$ very
quickly, $\rho _{xy}$ is insensitive to disorders and remains close to the
classical value at $W=4,5,6$($W_c=3.5$). The finite size effect of $
\rho _{xy}$ is shown in Fig. 2(b) at $M=16$. By increasing the sample length 
$L$ from $16$ to $64$, one sees that $\rho _{xy}$ converges to the dashed curve
(the classic value) very quickly whereas $\rho _{xx}$ keeps increasing
monotonically with the sample size. It is noted that both $\sigma _{xx}$ and 
$\sigma _{xy}$ calculated here are for square samples $L\times L$ using
Landauer\cite{land} and Kubo formula, respectively, and more than $2000$
disorder configurations are taken at $L=64$ and even more at smaller sample
sizes.

Such a classic behavior of $\rho_{xy}$ has been extensively observed
experimentally\cite{krav,song,hilke2} in weak magnetic field regime. 
$\rho_{xy}\propto 1/n_{\nu}$ in fact still holds at the critical point 
between the QH liquid and insulating regime I as previously shown 
experimentally\cite{song} and numerically.\cite{dns1} Since 
LLs are effectively coupled together at weak field and strong disorders, we 
believe that this phenomenon reflects the fact that the regime I is 
basically an Anderson insulator: $\rho_{xy}$ is always unrenormalized and 
remains at the classic value.\cite{lee} In other words, the 
insulator I in Fig. 1(c) should continuously evolve into the well-known 
Anderson insulator at zero magnetic field without changing the classic 
behavior of $\rho_{xy}$ while $\rho_{xx}$ is always divergent in the 
thermodynamic limit at zero temperature.

Now we consider the insulating regime II in Fig. 1(c). Along the scan line C in 
Fig. 1(c), the results of $\rho_{xx}$ and $\rho_{xy}$ are presented in 
Fig. 3(a). It shows that $\rho_{xy}$ remains at
quantized value $h/e^2$ while $\rho_{xx}$ arises almost an order of
magnitude from the critical value at $n_{\nu c}$ into the insulator
region. This is in contrast to the aforementioned classic behavior $
\rho_{xy}= \frac 1{n_{\nu}} h/e^2$  in the regime I.
Such a quantized $\rho_{xy}$ exists in the whole critical region
along the boundary between the $\nu=1$ QH state and the insulator. The
open circles in Fig. 1(c) at $W/W_c=1$ lies very close to the boundary of the 
two insulating regimes as indicated in our numerical calculations (how two
regimes exactly cross over will need a more careful study which is beyond the 
scope of the present paper). It is noted that in the transition region where 
$\rho_{xy}=h/e^2$ 
is observed, both $\sigma_{xx}$ and $\sigma_{xy}$ satisfy a one parameter 
scaling law\cite{huke}, which suggests that it is a consequence related to 
quantum phase transition.

The experimental results\cite{hilke2} of $\rho_{xx}$ and $\rho_{xy}$ are
presented for comparison in Fig. 3(b). It shows that the range of $n_{\nu}$ 
for the quantized $\rho_{xy}$ and the corresponding values of $\rho_{xx}$ are 
very close to our numerical ones. Here the temperature dependence of the
experimental data can be translated into the $L$-dependence of our numerical 
results at $T$=0 through a dephasing length $L_{in}$. To further 
compare with the experiments\cite{cole,diver}, the calculated $\rho_{xx}$ as a 
{\it scaling function} of the relative LL filling number 
$\delta n_{\nu}=n_{\nu}-n_{\nu c}$, i.e., $\rho_{xx}=f(\delta n_{L}/\nu_0)$, is 
shown in the inset of Fig. 3(a), where $\nu_0=c_0 (L/l_0)^{-1/x}$, $x=2.3$, 
and $l_0$ is the magnetic length ($c_0$ is a dimensionless constant). The 
experimental data (from Fig. 3 of Ref.\onlinecite{cole}) are also plotted in 
the inset using the $T$-dependent $\nu_0$ and an excellent agreement over a 
wide range of the scaling variable: $-2<\delta n_{\nu}/\nu_0<2$ is clearly 
shown.

Here we note that in some experiments\cite{shahar} whether the quantum
critical regime is reached is still controversial and there is an 
alternative explanation for the quantized $\rho_{xy}$ regime in which 
interactions may play\cite{the1} a crucial role for a non-scaling
behavior of the transport coefficients. An important distinction between 
such an interaction case and the present theory is that in the former case 
$\rho_{xy}$ is always well  quantized in the insulating regime while the 
quantization of $\rho_{xy}$ in Fig. 3(a) is mainly confined around the 
critical point $n_{\nu c}$ with $\rho _{xx}<10$ $h/e^2$ and $\rho_{xy}$ 
eventually will start to distinctly grow with $\rho_{xx}$ as $\rho_{xx}$ 
further increases\cite{dns2}.
Further experimental measurement in this regime may help to clarify this issue.

Finally, we would like to discuss a key physical distinction between the 
numerical phase diagram in Fig. 1(c) and the GPD in Fig. 1(a). In the latter
case, all the QH boundaries eventually float up to $n_{\nu}\rightarrow\infty$
at $B\rightarrow 0$ with the LL plateau structure in between remaining 
basically unchanged. But in both the TBM and the experiments, those vertical 
$\nu>1$ QH 
boundaries [see Fig. 1(c)] do not markedly ``float up'' in $n_{\nu}$ with 
increasing $W$ or reducing $B$ such that each LL plateau in between {\it 
never} floats away: only the width of the $\nu$-th QH plateau is reduced and 
vanishes eventually at the $\nu \rightarrow 0$ transition boundary. It then 
results in direct transitions and two insulating regimes in Fig. 1(c). 
To confirm this picture, we calculate the so-called equilibrium edge 
current\cite{pru1} which is proportional to 
$\partial n/\partial B|_{E_f}$\cite{exp} 
($n$ is the electron density and $E_f$ is the Fermi energy) and is 
$L$ independent.
The results (which are $B$-independent) are present in Fig. 4 
in which the peaks determine the centers 
of QH plateaus\cite{pru1}.  Indeed, such a quantity
is continuously reduced with increasing $W$ and eventually diminishes 
at $W_c$, but its peak positions at $W<W_c$ never move away which clearly 
indicates that the recovery of a Andersen insulator at strong disorder  in the 
integer QH system is due to the destruction of the plateaus instead of a 
float-up of the whole QH structure towards $n_{\nu}\rightarrow \infty$. 
  
To summarize, we have determined a numerical phase diagram of the integer QH 
state for the first time based on the TBM. The topology of such a phase 
diagram is remarkably similar to the experimental one obtained recently for 
the QH system. Two kinds of insulating regimes surrounding the QH plateau 
phase are identified whose transport properties, characterized by the classic 
and quantized values of the Hall resistance, respectively, are also in good 
agreement with the experiments. The nature of such a phase diagram can
be understood as a continuous narrowing and collapsing of the QH plateaus 
which are pinned around discrete  LL filling numbers without floating 
away.   

{\bf Acknowledgments} - The authors would like to acknowledge helpful
discussions with R. N. Bhatt, S. V. Kravchenko, X.-G. Wen, L. P. Pryadko,
A. Auerbach, and especially P. Coleridge and M. Hilke who also provided us
their experimental data prior to publication. This work is supported by the 
State of Texas through ARP Grant No. 3652707 and Texas Center for 
Superconductivity at University of Houston.

Fig. 1 The phase diagram in disorder - 
$1/n_{\nu}$ plane: (a) Theoretic globle phase 
diagram predicted in Ref.\onlinecite{phased}; (b) Experimental  one
in Ref. \cite{hilke1} (c) The present numerical result. 
Note that the scan line A in (c)  corresponds to  a
constant $V_G$ line in (b) (see text). 

Fig. 2 Hall resistance $\rho_{xy}$ (in units of $h/e^2$) as a function of 
$n_{\nu}$ along the scan line B in Fig. 1(c). The dashed curve 
represents the classic value of $1/n_{\nu}(h/e^2)=B/nec$. 
(a) $\rho_{xy}$ at different disorder strength $W$'s. The inset: $\rho_{xx}$
 vs. $n_{\nu}$. 
(b) The finite size effect of $\rho_{xy}$.

Fig. 3 (a) The longitudinal resistance $\rho_{xx}$ and Hall resistance
$\rho_{xy}$ (in units of $h/e^2$) versus $n_{\nu}$ along the scan line C  
in Fig. 1(c) at  $M=8$. The inset: The scaling function  
$\rho_{xx}=f(\delta n_{\nu}/\nu_0)$ 
obtained from the numerical calculation ($+$) and the experimental 
measurement (Fig. 3 of \cite{cole})($\bullet$). 
(b) Experimental data \cite{hilke2} at different 
temperatures (note that $\rho_{xx}$ is in units of $\rho_c=1.73 h/e^2$ 
according to Ref.\onlinecite{hilke2}).  

Fig. 4 The equilibrim edge current\cite{pru1} $\partial n/ \partial B|_{E_f}$
vs. $n_{\nu}$.   It indicates that the QH plateaus are pinned at 
integer ${n_{\nu} }'s$ until their destruction by disorder.


\begin{references}
\bibitem{phased} S. Kivelson, D. H. Lee and S. C. Zhang, Phys. Rev. B.
{\bf 46}, 2223(1992).
\bibitem{krav} S. V. Kravchenko et al., Phys. Rev. Lett. {\bf 75}, 910
(1995); A. A. Shashkin, G. V. Kravchenko, and V. T. Dolgopolov, JETP Lett.
{\bf 58}, 220 (1993); V. M. Pudalov et al., Sur. Sci. {\bf 305}, 107 (1994);
Physica B {\bf 194}, 1287 (1994).
\bibitem{song}  S. -H. Song et al., Phys. Rev. Lett. {\bf 78}, 2200 (1997);
D. Shahar et al., Phys. Rev. B ${\bf 52}$, R14372 (1995).
\bibitem{others}  C. H. Lee et al., Phys. Rev. B {\bf 58}, 10629 (1998).
\bibitem{hilke1} M. Hilke et al., preprint cond-mat/9906212.
\bibitem{qhe} For reviews see, The Quantum Hall Effect, edited by
R. E. Prange and S. M. Girvin (Springer-Verlag, New York, 1990).
\bibitem{laughlin} R. B. Laughlin, Phys. Rev. Lett. ${\bf 52}$, 2304 (1984);
D. E. Khmel'nitzkii,  Phys.  Lett. ${\bf 106A}$, 182 (1984).
\bibitem{hilke2}  M. Hilke et al., Nature ${\bf 395}$, 675(1998).
\bibitem{shahar} D. Shahar et al., Solid State Commun. ${\bf 107}$, 19 (1998);
D. Shahar et al., Phys. Rev. Lett. {\bf 79}, 479 (1997); D. Shahar et al.,
Science ${\bf 274}$, 589 (1996);
M. Hilke et al., Europhys. Lett. (1999); M. Hilke et al.,
Phys. Rev. B {\bf 56}, 15545 (1997).
\bibitem{dns} D. N. Sheng and Z. Y. Weng, Phys. Rev. Lett. ${\bf 78}$, 318
(1997).
\bibitem{xie} D. Z. Liu, et al., Phys. Rev. Lett. ${\bf 76}$, 975 (1996);
Phys. Rev. B. ${\bf 54}$, 4966 (1996).
\bibitem{others1}  H. Potempa et al., Physica B ${\bf 256}$, 591 (1998); Y.
Hatsugai, K. Ishibashi, and Y. Moritai, preprint cond-mat/9903223.
\bibitem{bhatt} K. Yang and R.N. Bhatt, Phys. Rev. B ${\bf 59}$, 8144 (1999);
Phys. Rev. Lett. ${\bf 76}$, 1316 (1996).
\bibitem{land} A. MacKinnon and B. Kramer, Z. Phys. {\bf 53},1 (1983);
 D. S. Fisher and P. A. Lee, Phys. Rev. B, ${\bf 23}$, 6851
(1981).
\bibitem{jiang}  I. Glozman et al., Phys. Rev. Lett. ${\bf 74}$, 594 (1995).
\bibitem{dns1}  D. N. Sheng and Z. Y. Weng, Phys. Rev. Lett. ${\bf 80}$,
580 (1998).
\bibitem{lee}  P. A. Lee and T. V. Ramakrishnan, Rev. Mod. Phys. {\bf 57}, 287
 (1985).
\bibitem{huke}  B. Huckestein and B. Kramer, Phys. Rev. Lett. ${\bf 64}$,
1437 (1990); B. Huckestein, Rev. Mod. Phys. {\bf 67}, 357 (1995).
\bibitem{cole} P. T. Coleridge and P. Zawadzki, preprint cond-mat/9903246.
\bibitem{diver}  R.T.F. van Schaijk et al., preprint cond-mat/9812035.
\bibitem{the1}  E. Shimshoni and A. Auerbach, Phys. Rev. B ${\bf 55}$, 9817
(1997); L. P. Pryadko and A. Auerbach, Phys. Rev. Lett. {\bf 82}, 1253 (1999).
\bibitem{dns2}  D. N. Sheng and Z. Y. Weng, Phys. Rev. B ${\bf 59}$, R7821
(1999).
\bibitem{pru1} A. M. M. Pruisken  in Ref. \cite{qhe}.
\bibitem{exp}  $\partial n/\partial B|_{E_f}=\int ^{E_f}
 (\rho_b (B+\Delta B)-\rho_b(B))dE/\Delta B$ ($\Delta B << B$) with 
$\rho_b $ as the bulk density of states,  see also Ref. \cite{pru1}.
\end{references}
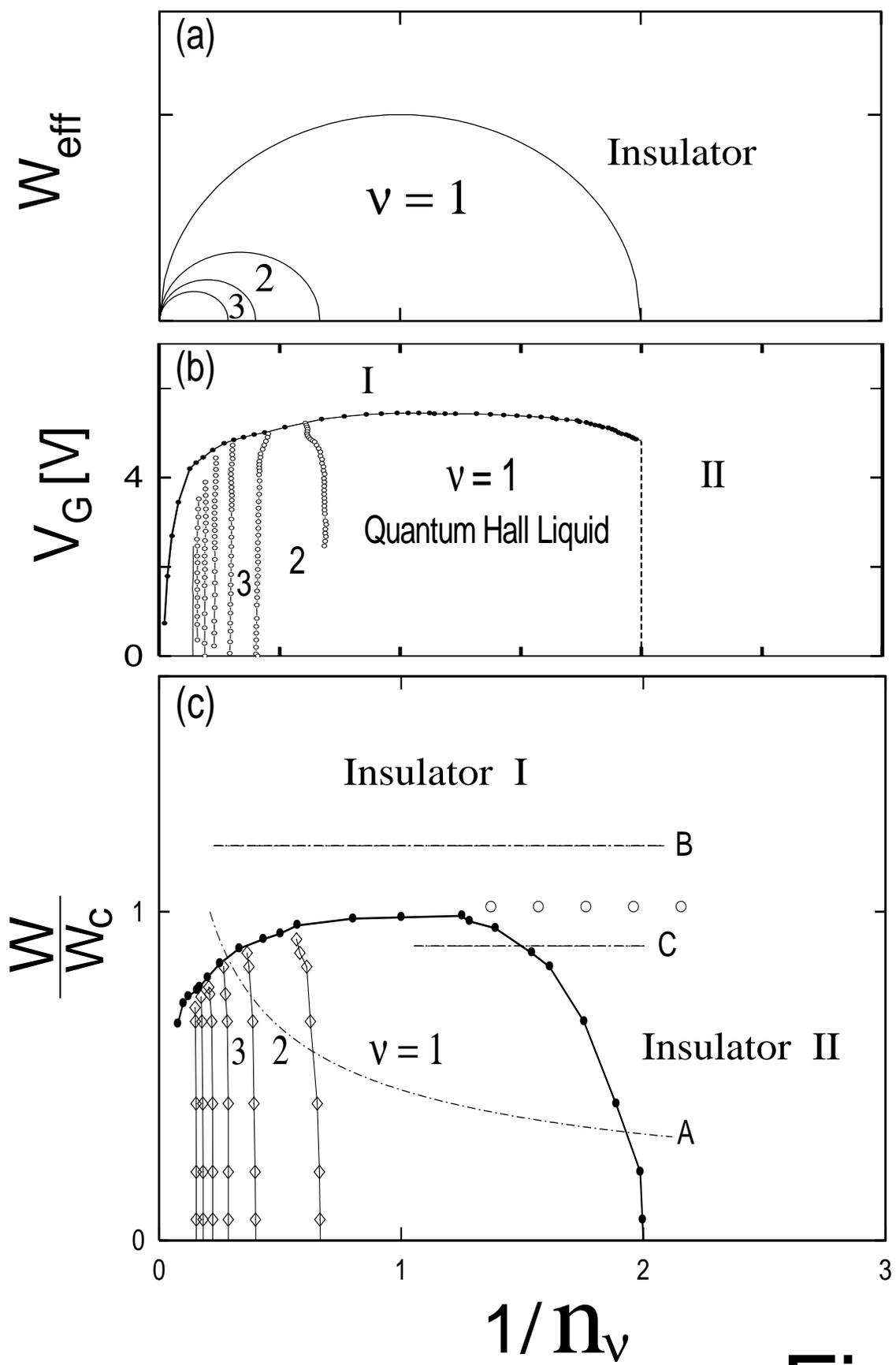
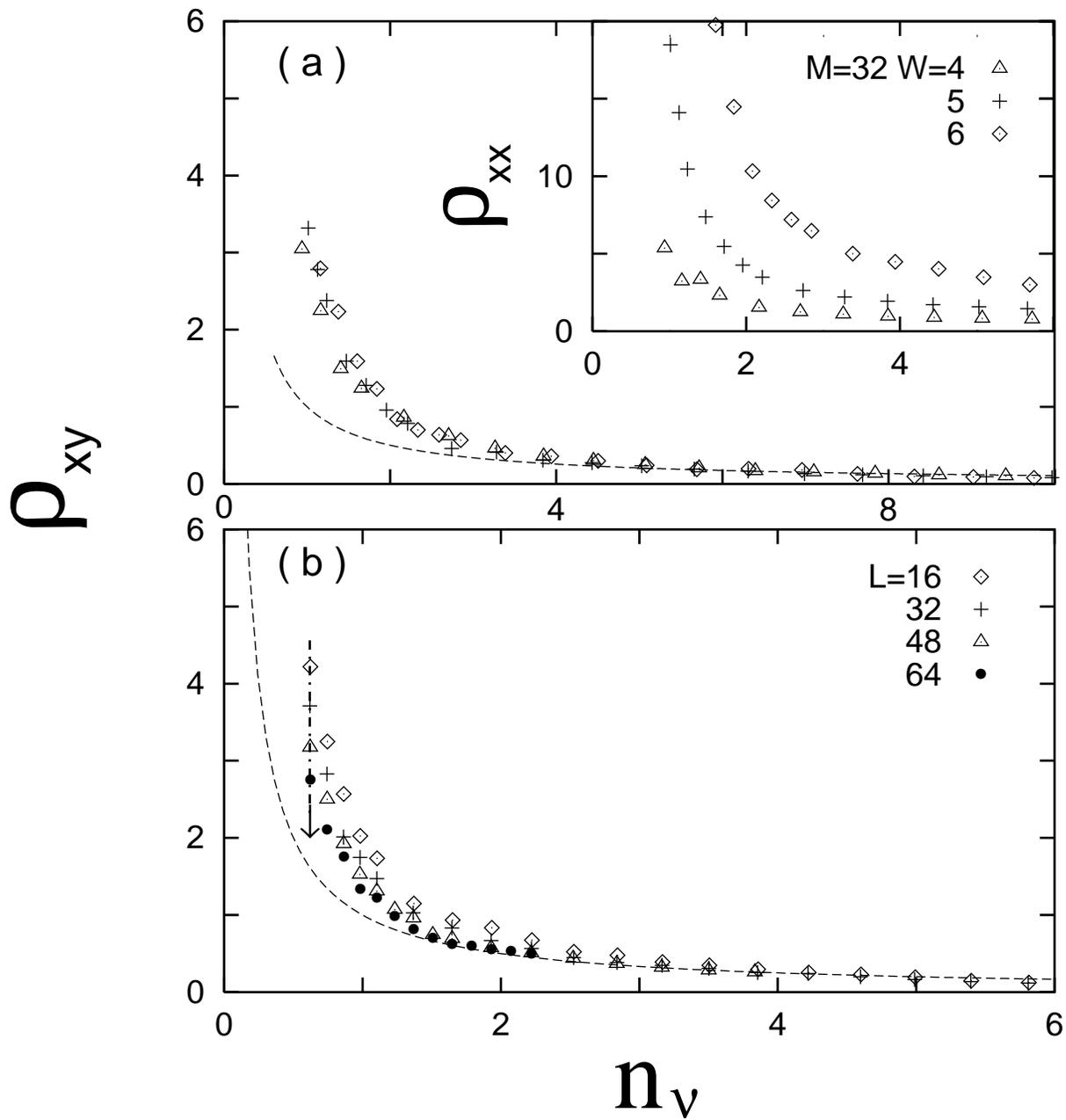
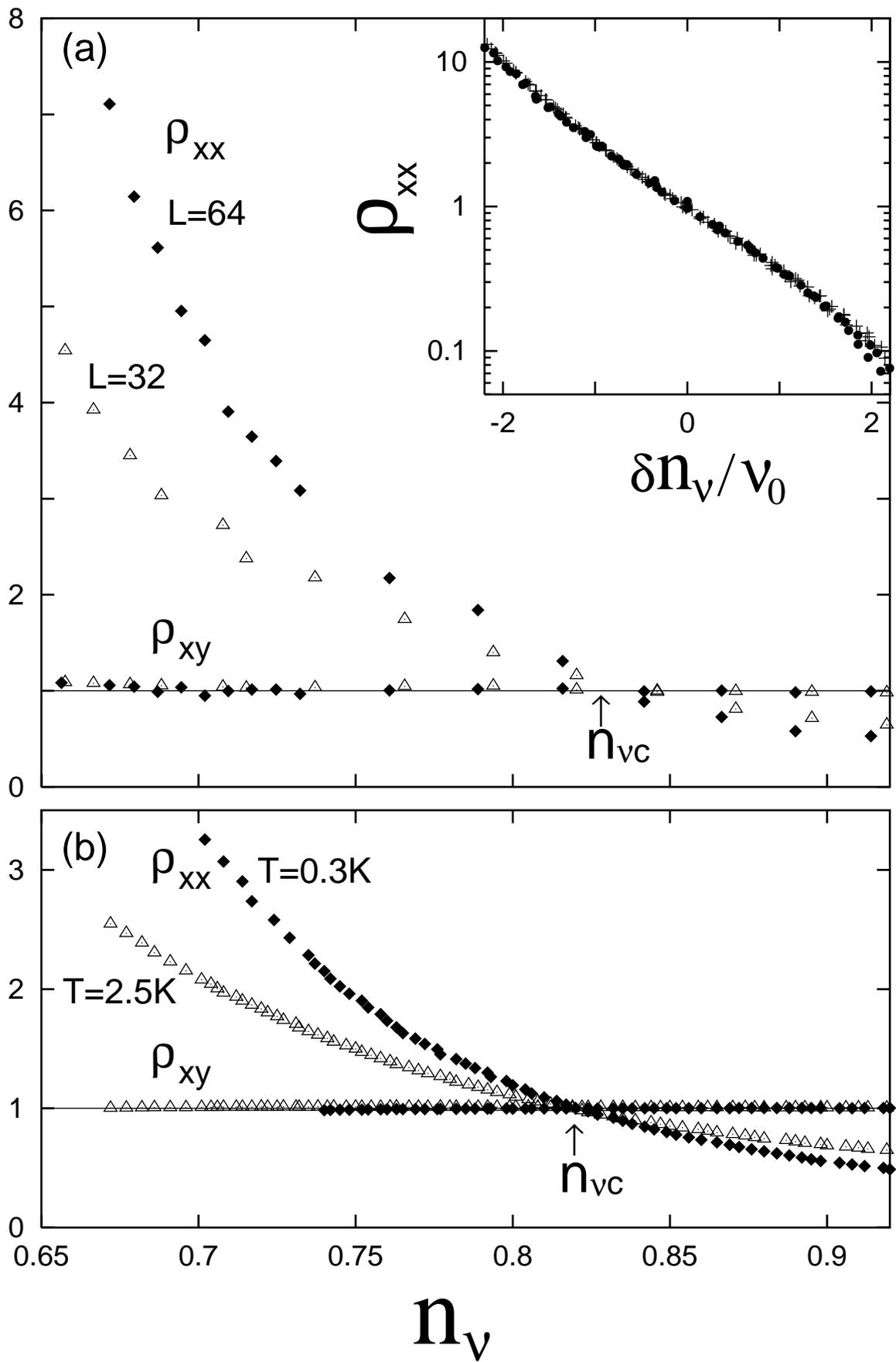
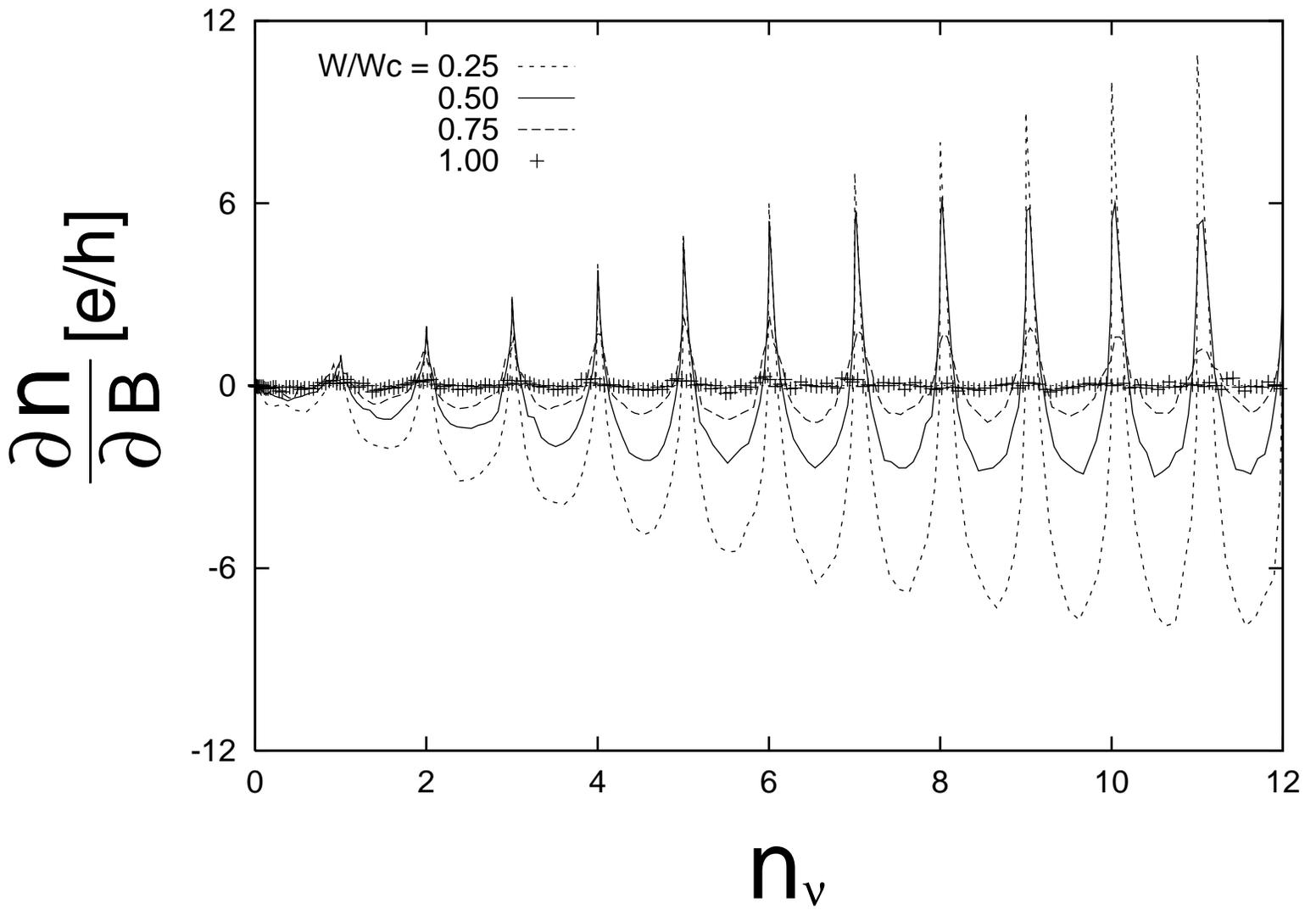
\end{document}